\documentclass[aip,rbi,reprint,graphicx]{revtex4-1} % for checking your page length
\usepackage[dvips]{graphicx}
\usepackage{color}

\begin{document}

%\preprint{}

\title{Cryogenic cooling with cryocooler on a rotating system} 


% \author{}
%\thanks{}
% \affiliation{}

\author{S.~Oguri}
\email[]{shugo@post.kek.jp}
\affiliation{Institute of Particle and Nuclear Studies, High Energy Accelerator Research Organization~(KEK), Oho, Tsukuba, Ibaraki 305-0801 Japan}
\author{J.~Choi}
\affiliation{Korea University, Anam-dong Seongbuk-gu, Seoul 136-713 Korea}
\author{M.~Kawai}
\affiliation{Institute of Particle and Nuclear Studies, High Energy Accelerator Research Organization~(KEK), Oho, Tsukuba, Ibaraki 305-0801 Japan}
\author{O.~Tajima}
% \email[]{osamu.tajima@kek.jp}
\affiliation{Institute of Particle and Nuclear Studies, High Energy Accelerator Research Organization~(KEK), Oho, Tsukuba, Ibaraki 305-0801 Japan}
\affiliation{Department of Particle and Nuclear Physics, School of High Energy Accelerator Science, The Graduate University for Advanced Studies (SOKENDAI), Shonan Village, Hayama, Kanagawa 240-0193 Japan}


\date{\today}

\begin{abstract}
We developed a system that continuously maintains a cryocooler for long periods on a rotating table.
A cryostat that holds the cryocooler is set on the table.
A compressor is located on the ground and supplies high-purity ($>$~99.999\%) and high-pressure (1.7~MPa) helium gas and electricity to the cryocooler.
The operation of the cryocooler and other instruments requires the development of interface components between the ground and rotating table.
A combination of access holes at the center of the table and two rotary joints allows simultaneous circulation of electricity and helium gas.
The developed system provides two innovative functions under the rotating condition;
cooling from room temperature and the maintenance of a cold condition for long periods.
We have confirmed these abilities as well as temperature stability
under a condition of continuous rotation at 20~revolutions per minute.
The developed system can be applied in various fields; 
e.g., in tests of Lorentz invariance, searches for axion, radio astronomy and cosmology, and application of radar systems.
In particular, there is a plan to use this system for a radio telescope observing cosmic microwave background radiation.
\end{abstract}


% \keywords{}

\maketitle %\maketitle must follow title, authors, abstract and \pacs

% If in two-column mode, this environment will change to single-column format so that long equations can be displayed. 
%\begin{widetext}
%$$\mbox{put long equation here}$$
%\end{widetext}

\section{Introduction}\label{intro}
Cooling of instruments on a rotating system is an important technology in science experiments; e.g., in tests of Lorentz invariance~\cite{PhysRevLett.95.040404, PhysRevA.71.050101}, searches for axion~\cite{PhysRevD.78.032006}, and when using radio telescopes for astronomy and cosmology~\cite{0004-637X-760-2-145}.
Thus far, experiments have been performed on either a continuous rotation system with a wet cooling (i.e., using refrigerant liquids such as liquid helium)~\cite{PhysRevLett.95.040404, PhysRevD.78.032006} or a system with periodic left--right motion through a limited angle and dry cooling (i.e., using cryocoolers)~\cite{PhysRevA.71.050101, 0004-637X-760-2-145}.
One of the greatest disadvantages of the former system is that the holding time of the refrigerant liquid limits the duration of the experiment.
The use of such a system requires frequent halts to the experiment and frequent refills of the liquid.
Therefore, it is difficult to perform an experiment for long periods under harsh conditions; e.g., at high altitude where human access is limited.
Merits of the latter system are that it continuously maintains a cold condition for long periods of time, and it does not require frequent maintenance.
However, its motion speed is limited. 
This sometimes reduces the experimental sensitivity.
For example, the pointing motion speed of a telescope limits the scan range because of a baseline fluctuation of instruments ($1/f$ noise); its importance is described in \cite{quiet_instruments}.

The combination of the merits of dry and wet coolers is a long-awaited technology.
We have developed such technology that continuously maintains a cold condition on a rotating system for long periods without frequent maintenance.
We are motivated to use the technology for a radio telescope observing the cosmic microwave background (CMB)~\cite{doi:10.1117/12.925816}.
In this paper, we describe the developed system and its performance.

\section{System maintaining dry cooling on a rotating table}
\begin{figure*}[htb]
\includegraphics[width=12.5cm]{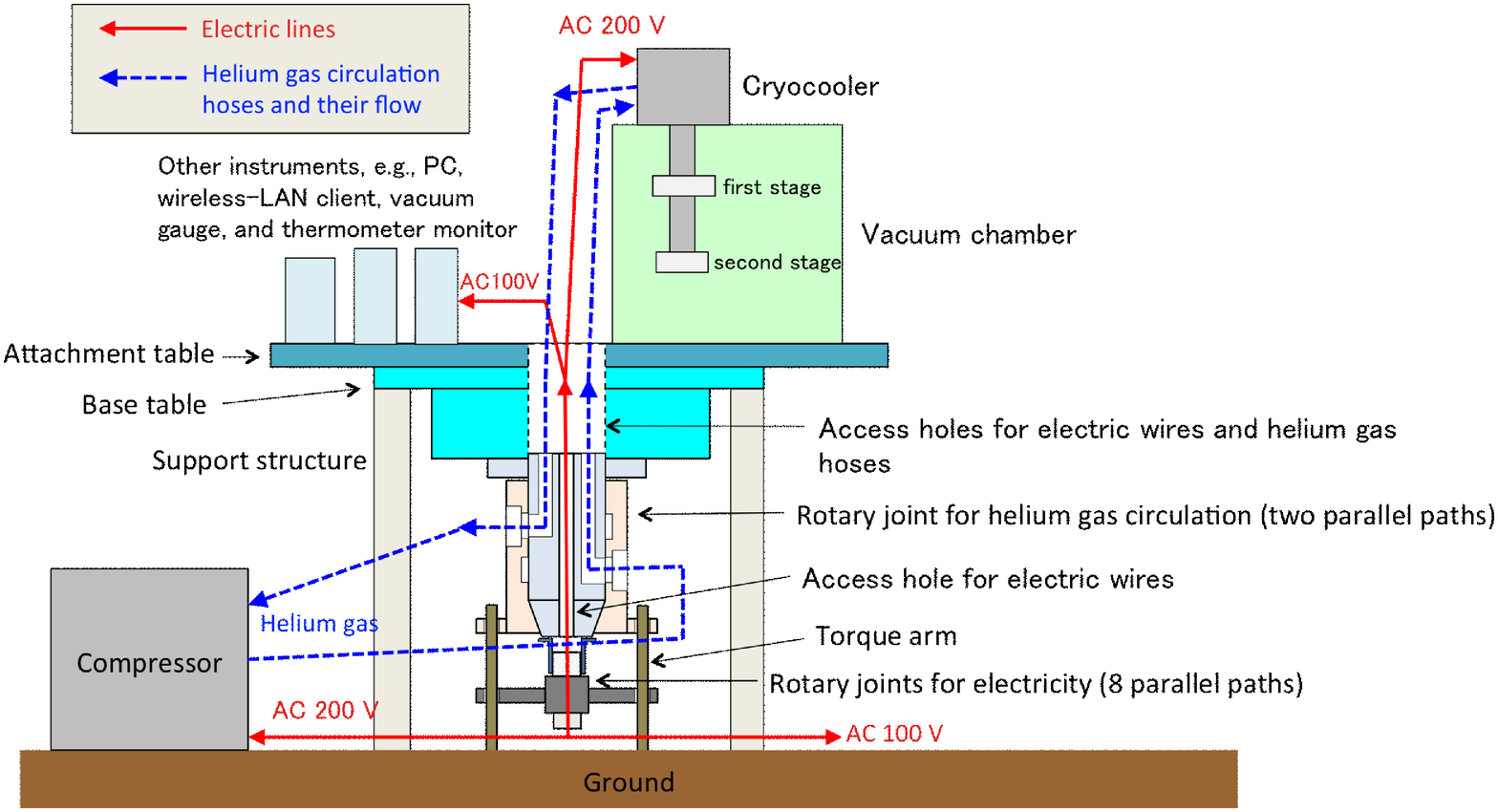}
\caption{Layout of the system maintaining dry cooling on a rotating table.
A cryostat (i.e., a vacuum chamber that holds a cryocooler) and other  instruments are set on the table.
Wireless LAN technology allows real-time communication with the instruments from the ground.
A custom-made rotary joint was developed for the helium gas transfer; there are two lines, a supply from and a return to a compressor located on the ground.
We also use a commercial rotary joint for the electricity.
Thus far we use seven electrical paths in parallel: three paths for the cryocooler control (three-phase AC 200~V), two for the other instruments (single-phase AC 100~V), and two for each ground.
Both the electrical wires and gas hoses are routed to the top of the table through a series of access holes at the center of the base and attachment tables.
\label{fig:overview_drawing}}
% \end{figure*}
%%%%%%%%%%%%%
%
%%%%%%%%%%%%%
% \begin{figure*}[htb]
\includegraphics[width=13.5cm]{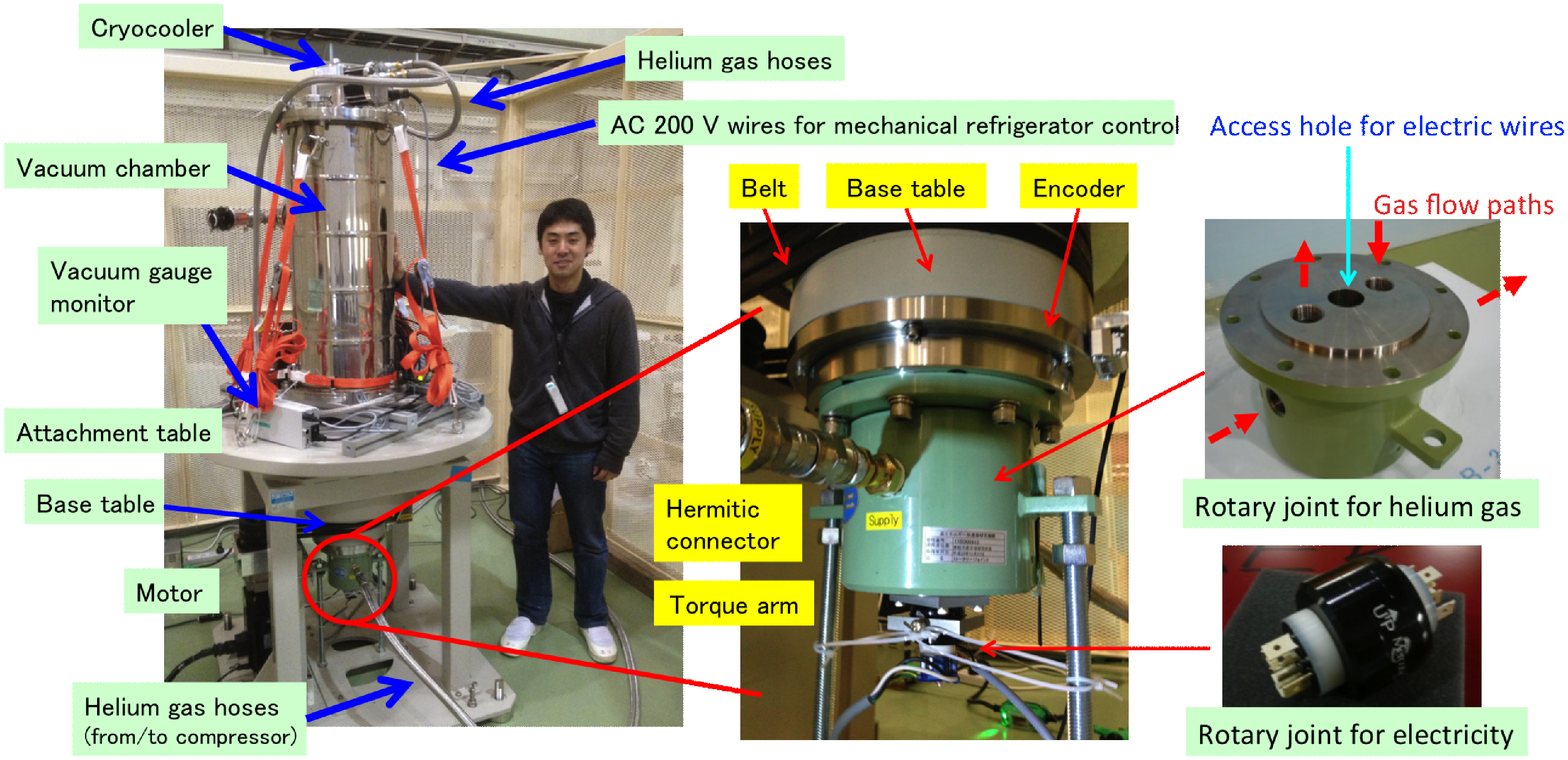}
\caption{Photograph of the system shown in Fig.~\ref{fig:overview_drawing} (left).
The compressor for the cryocooler is not shown.
A close-up of the interface between the rotating table and the ground; a series of two rotary joints connects to the bottom of the base table (middle).
Photographs of the rotary joints removed from the system (right).
\label{fig:overview_photo}}
\end{figure*}

Our system continuously maintains the operation of a cryocooler on a rotating table.
The layout of the system is shown in Fig.~\ref{fig:overview_drawing}, and photographs are shown in Fig.~\ref{fig:overview_photo}.
A cryostat that holds the cryocooler and other  instruments such as a personal computer (PC), temperature monitor, and vacuum gauge, are set on the rotating table.
Our unique design requires the development of interface components between the ground and rotating table.
Through a custom-made rotary joint, helium gas is transferred to the cryocooler from the compressor located on the ground.
We also use a commercial rotary joint for the electricity, which is needed to operate the cryocooler and other instruments.
Real-time communication with the instruments from the ground is easily achieved using a recent technology: a wireless local area network (wireless LAN).
%%%%%%%%%%%%%%
%%%%%%%%%%%%%%

\subsection{Rotating table}
The rotating table consists of a support structure, a base table, and an attachment table.
The shape of the base table is an upside-down top hat having a larger (smaller) diameter of 600~mm (257~mm) and height of 356~mm.
The base table is mounted on the support structure via an axial and thrust bearings.
A motor connected to a belt rotates the base table; the motor has sufficient torque to rotate a mass of 500~kg at 20 revolutions per minute~(rpm).
A rotary encoder and a series of two rotary joints are attached to the bottom of the base table.
We use a field-programmable gate array (FPGA) for encoder monitoring and motion control of the motor.
The attachment table (diameter of 1~m) is mounted on the top side of the base table,
and the cryostat and other instruments are set on it.
At the center of the base and attachment tables,
there is a series of large through holes, each with a diameters of 180~mm.
These are access holes for the hoses and wires passing through the rotary joints.

\subsection{Rotary joint for helium gas circulation}

For the circulation of helium gas,
a custom-made rotary joint was co-developed with Takeda Engineering Co. Ltd. in Japan.
The rotary joint has the shape of an upside-down top hat with a flange diameter of 205~mm, a body diameter of 159~mm, and height of 155~mm.
As shown in Fig.~\ref{fig:overview_drawing} and Fig.~\ref{fig:overview_photo}, there are two gas flow paths; a supply from and a return to the compressor located on the ground.
Circulation of the high-pressure gas is realized with a simple technique.
The rotary joint consists of two layers: an inner top-hat-shaped part and an outer cylinder.
The flange of the inner part is attached to the bottom-center position of the rotating table.
Therefore, the inner part rotates with the table.
Torque arms hold the outer cylinder, which does not rotate with the inner part.
The outer cylinder has two toroidal cavities on its inside surface, and each cavity has an access port (Rc3/4'') at the outside surface.
There are O-rings between the outer cylinder and the inner top hat that seal the cavities with each other.
There are two L-shaped gas lines (diameter of 3/4'') in the body of the inner part.
Two ports in the flange (which is parallel to the cylinder axis) are routes to the rotating table.
Two other ports on the opposite side always face the toroidal cavities of the outer cylinder allowing gas to flow through the two independent lines continuously.
Hermetic connectors (Eaton Co. Ltd., Aeroquip 5400 series) are attached at the input and output ports for each path.
The high-purity helium gas ($>$~99.999\%) is hermetically sealed in each path when the hoses are not joined.
To route the electric wires, there is an access hole (a through hole with diameter of 30~mm) at the center of the inner part.

%%%%%%%%%%%%%%
\begin{figure}[htb]
\includegraphics[width=8.5cm]{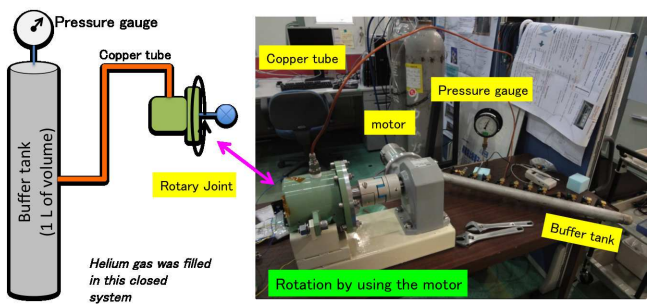}
\caption{Setup of a leak test of the gas rotary joint. 
The closed system was filled with high-pressure helium gas at 2.0~MPa;
a 1-L buffer tank and the input port of the joint are connected with a copper tube (diameter of 1/4'' and length of 1~m), and the output port of the joint is sealed with a ball valve.
A geared motor rotated the inner part of the joint.
We monitored the pressure while maintaining the rotation of the joint at 17~rpm.
We observed a zero-consistent leak rate in the test over eight days.
\label{fig:rj_leak_test}}
\end{figure}
%%%%%%%%%%%%%%
A leakage of gas would raise doubts about a joint.
Figure~\ref{fig:rj_leak_test} shows the setup of a leak test.
A buffer tank and the input port of the joint were connected with a copper tube.
The other side (i.e., the output port) of the joint was sealed with a ball valve.
We filled this closed system with helium gas.
The gas pressure was 2.0~MPa, which is an over-pressurized condition relative to the generic specification of the cryocooler's compressor (1.7~MPa).
The inner part of the joint was rotated using a geared motor at a constant speed of 17~rpm.
We found no leakage over eight days during the test; we obtained an upper limit for the leak rate of 2~cc/day.
The cryocooler and the compressor system have approximately a few hundred liters of buffer volume in general, and the gas pressure should be maintained within 2\%--3\% of the specification.
Therefore, the upper limit of the leak rate indicates that operation for more than a few years is possible.

\subsection{Rotary joint for electricity}
For the electricity, we use a commercial rotary joint, Model-830 manufactured by Mercotac Inc., which continuously maintains eight electrical paths in parallel.
A liquid metal (mercury) is used within the joint as an electrical conduction path, which is molecularly bonded to the contacts; therefore, the electrical connection has very low noise.
Thus far we have used three paths to control the cryocooler (three-phase AC~200~V),
two paths for other instruments (single-phase AC~100~V),
and two paths for their grounds.
Output wires are routed to the instruments on the table through a series of access holes at the center of the gas joint and table.

%%%%%%%%%%%%%%%%%%%%%%%%%%%%%%%

\section{Evaluation of System Performance}
The developed system provides two innovative functions under the rotating condition; cooling from room temperature, and maintenance of the cold condition for long-periods.
Temperature stability is also important in applications.
These three aspects of performance are investigated.

\subsection{Cooling from room temperature}
\begin{figure}[htb]
\includegraphics[width=4.0cm]{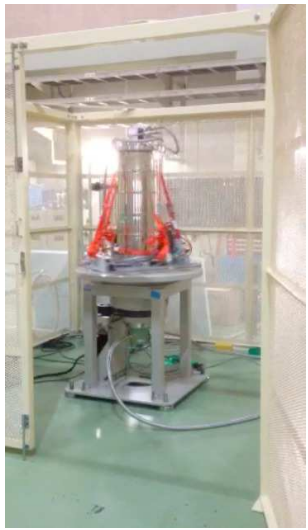}
\caption{
Movie of the cool-down test with rotation maintained at 20~rpm (enhanced online).
The GM cryocooler works on the rotating table.
\label{fig:movie}}
\end{figure}

\begin{figure}[htb]
\includegraphics[width=7.8cm]{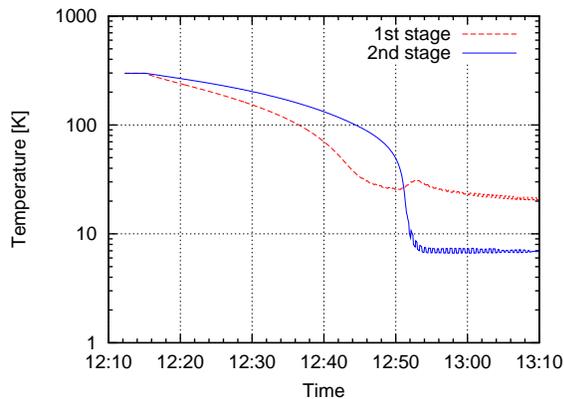}
\caption{
Time trends of temperatures at each cold stage of the GM cryocooler on the rotating system.
The specifications of the cryocooler, namely the obtained temperatures and the cool-down time from room temperature, were confirmed.
The motion of the displacer of the cryocooler produces a periodic temperature vibration with frequency of 1~Hz.
No temperature instability was found except for this effect.
\label{fig:trend_from_room_temp}}
\end{figure}

A cool-down test was performed under the condition of continuous rotation at 20~rpm.
For the purpose of thermal insulation, the cryostat was in a low-pressure ($\approx$0.01~Pa) condition in advance of the test.
We use a two-stage Gifford--McMahon (GM) cryocooler (Sumitomo Heavy Industries Ltd., RDK-408S) that has the ability to reach $\sim$23~K (in the first stage) and $\sim$7~K (in the second stage) without any additional thermal load.
The geometry of the vacuum chamber is diameter of 318~mm and height of 800~mm.
For simplicity of the test, we did not insert any additional material except for two silicon diode thermometers (Lakeshore Co. Ltd., DT-670), which monitor the temperatures of the two cold stages.
A movie of the test is shown in Fig.~\ref{fig:movie} (enhanced online). 
%%%%%%%%%%%%%%
%%%%%%%%%%%%%%
Figure~\ref{fig:trend_from_room_temp} shows time trends of temperatures in the two stages.
After one hour of cooling from room temperature, the system reached 21~K and 7~K in the first and second stages, respectively.
These temperatures are consistent with the specifications of the cryocooler.
The same temperatures were obtained without rotation of the table.
We thus have confirmed the ability of our system to provide a cold condition from room temperature while maintaining rotation.
%%%%%%%%%%%%%%
%%%%%%%%%%%%%%

\subsection{Temperature stability}\label{sec:temp_stability}
\begin{figure}[htb]
\includegraphics[width=8.2cm]{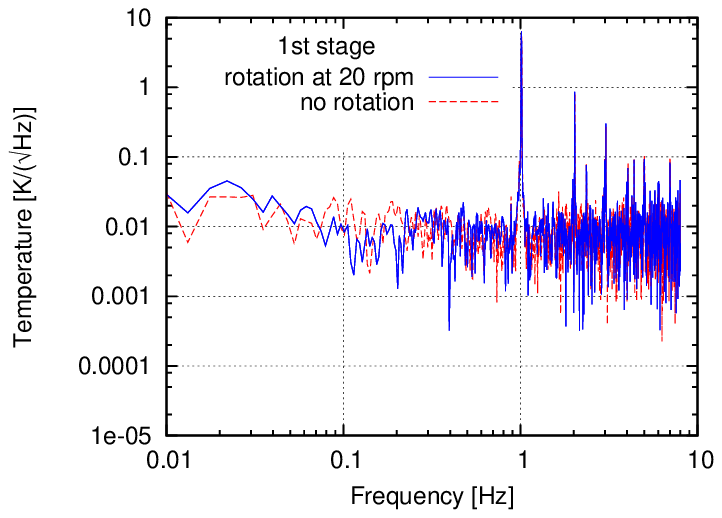}
\includegraphics[width=8.2cm]{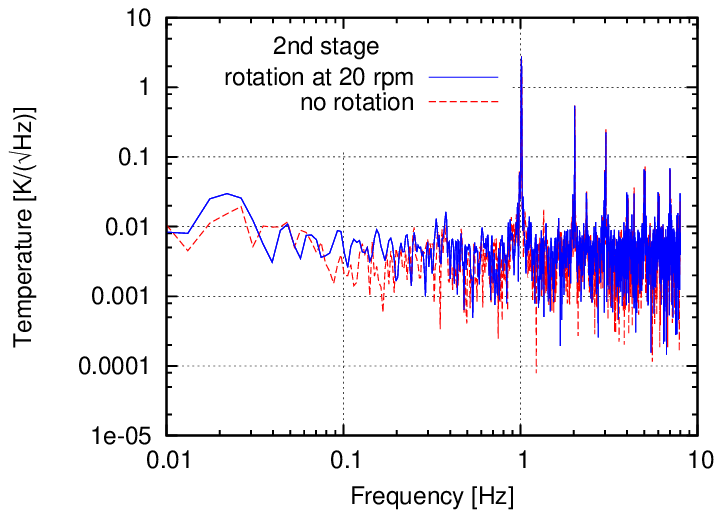}
\caption{Spectra of each the cold-head's temperature in Fourier space: first stage (top panel) and second stage (bottom panel).
The line style (solid or dashed) indicates the condition of table rotation.
The motion of the displacer in the GM cryocooler creates a cyclic temperature oscillation at 1~Hz.
This creates the lines in the spectrum at 1~Hz, 2~Hz, and so on.
No difference was found between the spectra with and without rotation of the table.
In particular, there is no characteristic feature around the rotation cycle at 20~rpm; i.e., 0.33~Hz.
We thus found no instability associated with rotation.
\label{fig:fft16sps}}
\end{figure}

A stable temperature under the rotating condition is required.
In particular, it is important to confirm that the table rotation does not produce any instability.
This is done by analyzing the Fourier space using the same setup as in the cool-down test.
We did not find any structure associated with the rotation cycle, as shown in  Fig.~\ref{fig:fft16sps}.
We also compared the spectra with and without rotation; there were no observable differences between the spectra.
We conclude that the rotation does not produce any instability above the level of the thermometer sensitivity of $\sim$0.01~K, which is almost three orders of magnitude lower than the temperature vibration caused by the displacer motion of the cryocooler.
%%%%%%%%%%%%%%
%%%%%%%%%%%%%%

\subsection{Long-term operation}
\begin{figure}[htb]
\includegraphics[width=8.7cm]{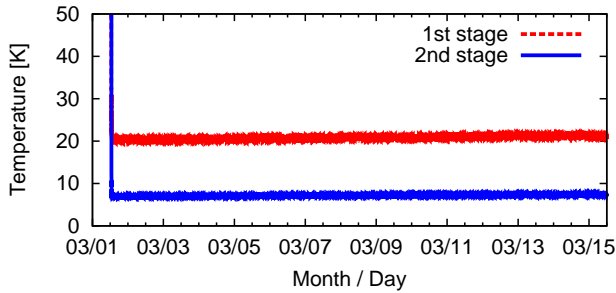}
\caption{Long-term trends of temperatures in each cold stage.
We continuously maintained the operation of the cryocooler and table rotation.
We cooled down from room temperature; Fig.~\ref{fig:trend_from_room_temp} presents the one-hour trends at the beginning of this plot.
\label{fig:temp_long}
}
\end{figure}

We checked stability during long-term operation.
Using the same setup as used in the previous tests, we measured time trends of temperatures as shown in Fig.~\ref{fig:temp_long}.
We continuously maintained the operation of the cryocooler and the rotation of the table.
The system properly maintained the cold condition for more than two weeks without any maintenance.
%%%%%%%%%%%%%%
%%%%%%%%%%%%%%

\begin{figure}[htb]
\includegraphics[width=8.7cm]{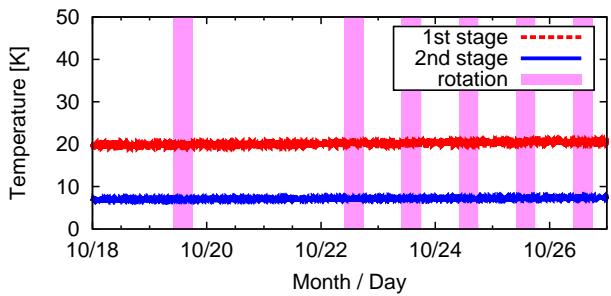}
\caption{
Time trends of temperatures in each cold stage when changing the rotating condition.
We repeated the start and stop sequence of rotation.
Shaded areas correspond to durations of the table rotation at 20~rpm.
Neither the starting nor stopping of rotation produced any instability.
\label{fig:temp_on_off}
}
\end{figure}

We compared the cooling conditions before and after table rotation.
We repeatedly started and stopped the rotation as shown in Fig.~\ref{fig:temp_on_off}; this is a simulation of the frequent maintenance of the rotating table or instruments.
% No instability associated with the start and stop sequence was found.
No temperature variation associated with the start and stop sequence was found.
Zero-consistent leakage of the helium gas (upper bound of 0.05~MPa) was observed during the test.
%%%%%%%%%%%%%%
%%%%%%%%%%%%%%

%%%%%%%%%%%%%%%%%%%%%%%%%%%%%%%

\section{Conclusion}

We developed a system that continuously maintains a cold condition on a rotating table for long periods.
A cryocooler (i.e., dry cooler) is held on a cryostat set on the rotating table.
A compressor is located on the ground and supplies high-purity ($>$~99.999\%) and high-pressure (1.7~MPa) helium gas and electricity to the cryocooler. 
Additional electricity has to be supplied to other instruments on the rotating table.
The combination of two rotary joints and a series of access holes at the center of the table allows the simultaneous transfer of helium gas and electricity.
While maintaining rotation of the table, we were able to cool the cryostat from room temperature to the specification temperature of the GM cryocooler.
We confirmed temperature stability; there is no temperature instability associated with the cycle of the table rotation.
Moreover, neither the starting nor stopping of rotation affected the cooling condition.
We also confirmed the ability of the continuous operation for long periods without any maintenance; the system maintained the cold condition for more than two weeks.

The system can be extended to manage lower temperatures using the higher-power cryocoolers or a combination of cryocoolers.
Various applications in scientific fields are expected; e.g., tests of Lorentz invariance, axion searches, and the use of radio telescopes to observe the CMB.
In particular, a experiment to observe the CMB polarization plans to use this technology~\cite{doi:10.1117/12.925816}.
Application to systems used in other fields, such as a radio detection and ranging (radar) system for meteorological and military purposes, is also expected.

\begin{acknowledgments}
This work is supported by Grants-in-Aid for Scientific Research from The Ministry of Education, Culture, Sports, Science, and Technology, Japan (KAKENHI 23684017 and  21111003).
It is also partially supported by the Center for the Promotion of Integrated Sciences (CPIS) of SOKENDAI.
We are grateful for the cooperation of Takeda Engineering Co. Ltd., Bee
 Beans Technologies Co. Ltd., and G-tech Co. Ltd.
We also acknowledge Koji Ishidoshiro, Masaya Hasegawa, Chiko Otani, and Masa Fukumuro for their useful comments about the applications of the developed system.
\end{acknowledgments}

%\bibliography{ref.bib}
% Run this once to generate your BBL file. Then copy the contents of your BBL file into your main latex file, commenting out "\bibliography"

\end{document}